\documentstyle[aps,prl,epsf]{revtex}

\begin{document}
\draft \twocolumn[\hsize\textwidth\columnwidth\hsize\csname
@twocolumnfalse\endcsname

\preprint{}

\title{Chaotic inflation on the brane}

\author{Roy Maartens, David Wands, Bruce A. Bassett and Imogen P.C.
Heard}

\address{Relativity and Cosmology Group, School of Computer Science
and Mathematics,\\University of Portsmouth, Portsmouth~PO1~2EG,
Britain}

%\date{\today}
\maketitle
\begin{abstract}
We consider slow-roll inflation in the context of recently
proposed four-dimensional effective gravity induced on the
world-volume of a three-brane in five-dimensional Einstein
gravity. We find significant modifications of the simplest chaotic
inflationary scenario when the five-dimensional Planck scale is
below about $10^{17}$~GeV. We use the comoving curvature
perturbation, which remains constant on super-Hubble scales, in
order to calculate the spectrum of adiabatic density perturbations
generated.  Modifications to the Friedmann constraint equation
lead to a faster Hubble expansion at high energies and a more
strongly damped evolution of the scalar field. This assists
slow-roll, enhances the amount of inflation obtained in any given
model, and drives the perturbations towards an exactly
scale-invariant Harrison-Zel'dovich spectrum. In chaotic inflation
driven by a massive scalar field we show that inflation can occur
at field values far below the four-dimensional Planck scale,
though above the five-dimensional fundamental scale.
\end{abstract}
\vskip 1pc \pacs{98.80.Cq \ \ 04.50.+h
\hspace*{3cm}PU-RCG-99/24
\hspace*{3cm}hep-ph/9912464}
]

\section{Introduction}

There is considerable interest in higher dimensional cosmological
models motivated by superstring theory solutions where matter
fields (related to open string modes) live on a lower dimensional
brane while gravity (closed string modes) can propagate in the
bulk~\cite{general}. In such a scenario the extra dimension need
not be small~\cite{gia1}, and may even be infinite if non-trivial
geometry can lead gravity to be bound to the three-dimensional
subspace on which we live at low
energies~\cite{Visser85,RanSun99,cg,manyinfd}. One possibility of
great importance arising from these ideas is the notion that the
fundamental Planck scale $M_{4+d}$ in $4+d$ dimensions can be
considerably smaller than the effective Planck scale,
$M_4=1.2\times 10^{19}$~GeV, in our four-dimensional spacetime,
which would have profound consequences for models of the very
early universe.

In this paper we investigate the impact of such a scenario when
$d=1$~\cite{RanSun99} for simple chaotic inflation models.
Specific models of inflation have previously been discussed with
finite compactified dimensions, scalar fields in the bulk and/or
multiple branes (see, e.g.,~\cite{kl,gia2,cgs}). Our aim is to
quantify the minimal modification of slow-roll inflation in the
brane scenario for arbitrary inflaton potentials on the brane,
independent of the dynamics of the bulk, while assuming stability
of the brane. If Einstein's equations hold in the five-dimensional
bulk, with a cosmological constant as source, and the matter
fields are confined to the 3-brane, then Shiromizu et
al.~\cite{ShiMaeSas99} have shown that the four-dimensional
Einstein equations induced on the brane can be written as
\begin{equation}
\label{Einstein} G_{\mu\nu} = -\Lambda_4 g_{\mu\nu} +
\left({8\pi\over M_{4}^2}\right) T_{\mu\nu}  + \left({8\pi\over
M_5^3}\right)^2 \pi_{\mu\nu} - E_{\mu\nu}\,,
\end{equation}
where $T_{\mu\nu}$ is the energy-momentum tensor of matter on the
brane, $\pi_{\mu\nu}$ is a tensor quadratic in $T_{\mu\nu}$, and
$E_{\mu\nu}$ is a projection of the five-dimensional Weyl tensor,
describing the effect of bulk graviton degrees of freedom on
brane dynamics. The effective cosmological constant $\Lambda_4$ on
the
brane is determined by the five-dimensional bulk cosmological
constant $\Lambda$ and the 3-brane tension $\lambda$ as
\begin{equation}
\Lambda_4= {4\pi \over M_5^3} \left( \Lambda + {4\pi\over3M_5^3}
\lambda^2 \right) \,,\label{2}
\end{equation}
and the four-dimensional Planck scale is given by
\begin{equation}
M_4 = \sqrt{{3\over 4\pi}} \left( {M_5^2\over\sqrt{\lambda}}
\right) M_5 \,.\label{3}
\end{equation}

In a cosmological scenario in which the metric projected onto the
brane is a spatially flat Friedmann-Robertson-Walker model, with
scale factor $a(t)$, the Friedmann equation on the brane has the
generalized form~\cite{bdel}
\begin{equation}
\label{Friedmann} H^2={\Lambda_4\over3} +
\left({8\pi \over 3M_4^2}\right) \rho +
\left({4\pi\over 3M_5^3}\right)^2\rho^2+{{\cal E} \over
a^4} \,,
\end{equation}
where ${\cal E}$ is an integration constant arising from
$E_{\mu\nu}$, and thus transmitting bulk graviton influence onto
the brane. This term appears as a form of ``dark
radiation"~\cite{bdel,dr} affecting primordial nucleosynthesis and
the heights of the acoustic peaks in the cosmic microwave
background radiation, because it is decoupled from matter on the
brane and behaves like an additional collisionless (and isotropic)
massless component.  Thus observations can be used to place limits
on $|{\cal E}|$.  However, during inflation this term will be
rapidly diluted, and we can neglect it. We will also assume that
the bulk cosmological constant
$\Lambda\approx-4\pi\lambda^2/3M_5^3$ so that $\Lambda_4$ is
negligible, at least in the early universe. This fine-tuning is
the restatement in the brane-world scenario of the cosmological
constant problem and we do not attempt to solve it here.

The crucial correction in what follows is the term quadratic in
the density, which modifies the expansion dynamics at densities
$\rho\gtrsim\lambda$. This can be seen on rewriting
Eq.~(\ref{Friedmann}) using Eq.~(\ref{3}), when $\Lambda_4=0$ and
${\cal E}=0$, to give
\begin{equation}
\label{Friedmann'} H^2={8\pi\over 3M_4^2}\rho\left[1+ {\rho\over
2\lambda}\right] \,.
\end{equation}
Note that in the limit $\lambda\to\infty$ we recover standard
four-dimensional general relativistic results (neglecting ${\cal
E}$). The quadratic modification will dominate at high energies
for moderate $\lambda$, but must be sub-dominant at
nucleosynthesis. Since it decays as $a^{-8}$ during the radiation
era, it will rapidly become negligible thereafter. The
nucleosynthesis limit implies that $\lambda\gtrsim (1\mbox{
MeV})^4$, and by Eq.~(\ref{3}) this gives~\cite{cgs}
\begin{equation}
M_5 \gtrsim \left({1 \mbox{ MeV}\over M_4}\right)^{2/3}M_4\sim 10\
{\rm TeV} \,. \label{4}
\end{equation}
A more stringent constraint may be obtained if the fifth dimension
is infinite, by requiring that relative corrections to the
Newtonian law of gravity, which are of order
$M_5^6\lambda^{-2}r^{-2}$ (see, e.g.,~\cite{RanSun99}), should be
small on scales $r\gtrsim 1$~mm. Using Eq.~(\ref{3}), this gives
$M_5>10^5$~TeV.

\section{Slow-roll inflation on the brane}

We will consider the case where the energy-momentum tensor
$T_{\mu\nu}$ on the brane is dominated by a scalar field $\phi$
(confined to the brane) with self-interaction potential $V(\phi)$.
The field satisfies the Klein-Gordon equation
\begin{equation}
\ddot{\phi}+3H\dot{\phi}+V'(\phi)=0\,, \label{5}
\end{equation}
since $\nabla^\nu T_{\mu\nu}=0$ on the brane. In four-dimensional
general relativity, the condition for inflation is
$\dot{\phi}^2<V(\phi)$, i.e., $p<-{1\over3}\rho$, where
$\rho={1\over2}\dot{\phi}^2+V$ and $p={1\over2}\dot{\phi}^2-V$.
This guarantees $\ddot{a}>0$. The modified Friedmann equation
leads to a {\em stronger condition for inflation:} using
Eqs.~(\ref{Friedmann'}) and (\ref{5}), we find that
\begin{equation}
\ddot{a}>0~~\Rightarrow~~p<-\left[{\lambda+2\rho \over \lambda
+\rho}\right]{\rho \over 3}\,. \label{6}
\end{equation}
As $\lambda\rightarrow\infty$, this reduces to the violation of
the strong energy condition, but for $\rho>\lambda$, a more
stringent condition on $p$ is required for accelerating expansion.
In the limit $\rho/\lambda\rightarrow\infty$, we have
$p<-{2\over3}\rho$. When the only matter in the universe is a
self-interacting scalar field, the condition for inflation becomes
\begin{equation}
\label{endinf}
\dot\phi^2 - V + {\dot\phi^2 + 2V \over 8\lambda}(5\dot\phi^2-2V) < 0
\,,
\end{equation}
which reduces to $\dot{\phi}^2<V(\phi)$ when
$(\dot\phi^2+2V)\ll\lambda$.

Assuming that the ``brane energy condition" in Eq.~(\ref{6}) is
satisfied, we now discuss the dynamics of the last 50 or so
e-foldings of inflation. Within the the slow-roll approximation,
we assume that the energy density is dominated by the
self-interaction energy of the scalar field and that the scalar
field evolution is strongly damped, which implies
\begin{eqnarray}
H^2 &\simeq&  \left({8\pi \over 3M_{4}^2}\right)V\left[
1+{V\over2\lambda} \right]\,,
 \label{7}\\
\dot\phi &\simeq & -{V'\over 3H}\,, \label{8}
\end{eqnarray}
where we use `$\simeq$' to denote equality within the slow-roll
approximation. The term in square brackets is the
brane-modification to the standard slow-roll expression for the
Hubble rate. For $V\gg \lambda$, Eqs. (\ref{3}) and (\ref{7}) give
$H\simeq (4\pi/3)V/M_5^3$ consistent with the ``non-linear''
regime discussed in Ref.~\cite{kl}.

Requiring the slow-roll approximation to remain consistent with the
full evolution equations places constraints on the slope and
curvature
of the potential. We can define two slow-roll parameters
\begin{eqnarray}
\label{epsilon}
\epsilon &\equiv&{M_{4}^2 \over 16\pi} \left( {V' \over V}
\right)^2 \left[ {2\lambda(2\lambda+2V)\over(2\lambda+V)^2}
\right]  \,,\label{10}\\
\label{eta}
\eta &\equiv& {M_{4}^2 \over 8\pi} \left(
{V'' \over V} \right) \left[ {2\lambda \over 2\lambda+V} \right]
\,.\label{11}
\end{eqnarray}
Self-consistency of the slow-roll approximation then requires
${\rm max}\{\epsilon,|\eta|\}\ll1$. At low energies,
$V\ll\lambda$, the slow-roll parameters reduce to the standard
form (see, e.g.,~Refs.\cite{LidLyt93,LidLidKol97}). However at
high energies, $V\gg\lambda$, the extra contribution to the Hubble
expansion helps damp the rolling of the scalar field and the new
factors in square brackets become $\approx\lambda/V$. Thus {\em
brane effects ease the condition for slow-roll inflation for a
given potential.}

The number of e-folds during inflation is given by $N =
\int_{t_{\rm i}}^{t_{\rm f}} Hdt$, which in the slow-roll
approximation becomes
\begin{equation}
\label{efold}
N \simeq - {8\pi  \over M_{4}^2}\int_{\phi_{\rm i}}^{\phi_{\rm
f}}{V\over V'} \left[ 1+{V \over 2\lambda}
 \right]  d\phi \,.\label{12}
\end{equation}
The effect of the modified Friedmann equation at high energies is
to increase the rate of expansion by a factor $[V/2\lambda]$,
yielding more inflation between any two values of $\phi$ for a
given potential. Thus we can obtain a given number of e-folds for
a {\em smaller} initial inflaton value $\phi_{\rm i}$. For
$V\gg\lambda$, Eq. (\ref{12}) becomes $N \simeq - (128\pi^3/3
M_{5}^6)\int_{{\rm i}}^{{\rm f}}(V^2/ V')d\phi$.

\section{Perturbations on the brane}

The key test of any inflation model, or any modified gravity
theory during inflation, will be the spectrum of perturbations
produced due to quantum fluctuations of the fields about their
homogeneous background values. To date there has been no study of
linear perturbations about a four-dimensional
Friedmann-Robertson-Walker universe on the brane for the modified
four-dimensional Einstein equations given in Eq.~(\ref{Einstein}).
The key uncertainty here comes from the tensor $E_{\mu\nu}$, which
describes the effect of tidal forces and gravitational waves in
the vacuum five-dimensional bulk and whose evolution is not
completely determined by the four-dimensional effective theory
alone. In what follows we set $E_{\mu\nu}=0$, effectively
neglecting back-reaction due to metric perturbations in the fifth
dimension. This is consistent with a homogeneous density of matter
on the brane~\cite{ShiMaeSas99} and thus is valid even in the
presence of scalar field perturbations in the slow-roll limit
(where $V'\to0$), but we note that a full investigation is
required to discover when back-reaction will have a significant
effect.

To quantify the amplitude of scalar (density) perturbations we
evaluate the gauge-invariant quantity~\cite{BarSteTur83}
\begin{equation}
\label{defzeta}
\zeta \equiv \psi-{H\over\dot\rho}\delta\rho \,,
\end{equation}
which reduces to the curvature perturbation, $\psi$, on uniform
density hypersurfaces where $\delta\rho=0$. The four-dimensional
energy-conservation equation, $\nabla^\nu T_{\mu\nu}=0$, for
linear perturbations (in an arbitrary gauge) on large scales,
requires that
\begin{equation}
\label{dotrho}
\delta\dot\rho + 3H(\delta\rho+\delta p) + 3(\rho+p) \dot\psi = 0
\,,
\end{equation}
where we have neglected spatial gradients. We can apply
Eq.~(\ref{dotrho}) on uniform density hypersurfaces, where
$\delta\rho=0$ and $\psi=\zeta$, [or, equivalently, use the
gauge-invariant definition of $\zeta$ given in Eq.(\ref{defzeta})]
to obtain
\begin{equation}
\dot\zeta = - H {\delta p_{\rm nad} \over \rho+p} \,.
\end{equation}
Hence $\zeta$ is conserved on large scales for purely adiabatic
perturbations, for which the non-adiabatic pressure perturbation,
$\delta p_{\rm nad}\equiv \delta p/\dot{p} - \delta\rho/\dot\rho$
vanishes. This gauge-invariant result is a consequence of the
local conservation of energy-momentum in four dimensions, and is
independent of the form of the gravitational field
equations~\cite{WMLL}.

The curvature perturbation on uniform density hypersurfaces
is given in terms of the scalar field fluctuations on spatially
flat hypersurfaces, $\delta\phi$, by
\begin{equation}
\zeta = {H\delta\phi\over\dot\phi} \,.
\label{9}
\end{equation}
The field fluctuations at Hubble crossing ($k=aH$) in the
slow-roll limit are given by
$\langle\delta\phi^2\rangle\simeq\left({H/2\pi} \right)^2$. Note
that this result for a massless field in de Sitter space is also
independent of the gravity theory~\cite{WMLL}. For a single scalar
field the perturbations are adiabatic and hence the curvature
perturbation $\zeta$ can be related to the density perturbations
when modes re-enter the Hubble scale during the matter dominated
era which is given (using the notation of Ref.~\cite{LidLidKol97})
by $A_{\sc s}^2 = 4\langle \zeta^2 \rangle/25$. Using the
slow-roll equations and Eq.~(\ref{9}), this gives
\begin{equation}
\label{AS}
A_{\sc s}^2 \simeq \left . \left({512\pi\over75 M_4^6}\right) {V^3
\over V^{\prime2}}\left[ {2\lambda + V \over 2\lambda} \right]^3
\right|_{k=aH}\,.\label{14}
\end{equation}
Thus the amplitude of scalar perturbations is {\em increased}
relative
to the standard result at a fixed value of $\phi$ for a given
potential.

The scale-dependence of the perturbations is described by the
spectral tilt
\begin{equation}
n_{\sc s}-1\equiv {d\ln A_{\sc s}^2 \over d\ln k} \simeq
-6\epsilon + 2\eta \,,\label{15}
\end{equation}
where the slow-roll parameters are given in Eqs.~(\ref{epsilon})
and~(\ref{eta}). Because these slow-roll parameters are both
suppressed by an extra factor $\lambda/V$ at high energies, we see
that the spectral index is {\em driven towards the
Harrison-Zel'dovich spectrum}, $n_{\sc s}\to1$, as
$V/\lambda\to\infty$.

The tensor (gravitational wave) perturbations are bound to the
brane at long-wavelengths~\cite{RanSun99} and decoupled from the
matter perturbations to first-order, so that the amplitude on
large scales is simply determined by the Hubble rate when each
mode leaves the Hubble scale during inflation. The amplitude of
tensor perturbations at Hubble crossing is given
by~\cite{LidLidKol97}
\begin{equation}
A_{\sc t}^2 = {4\over 25\pi} \left.
\left( {H\over M_{4}} \right)^2\right|_{k=aH}
\,.\label{16}
\end{equation}
In the slow-roll approximation this yields
\begin{equation}
A_{\sc t}^2 \simeq {32 \over 75M_{4}^4}V \left. \left[ {2\lambda + V
\over
2\lambda} \right] \right|_{k=aH} \,.\label{17}
\end{equation}
Again, the tensor amplitude is {\em increased} by brane effects,
but by a smaller factor than the scalar perturbations. The tensor
spectral tilt is
\begin{equation}
n_{\sc t}\equiv {d\ln A_{\sc t}^2 \over d\ln k} \simeq -2\epsilon
\,,\label{18}
\end{equation}
so that the ratio between the amplitude of tensor and scalar
perturbations is given by
\begin{equation}
{A_{\sc t}^2 \over A_{\sc s}^2} \simeq \epsilon \left. \left[
{\lambda
\over \lambda+V} \right]\right|_{k=aH}  \,.\label{19}
\end{equation}
Thus the standard observational test for consistency
condition~\cite{LidLidKol97} between this ratio and the tilt of
the gravitational wave spectrum is modified by the pre-factor
$\lambda/(\lambda+V)$, which becomes small at high energies.
Although the amplitude of both tensor and scalar perturbations is
enhanced due to the increased Hubble rate, the overall effect is
to suppress the contribution of tensor perturbations relative to
the scalar modes for a given potential $V$.

\section{A simple model}

As an example we investigate the simplest chaotic inflation model
driven by a scalar field with potential
$V={1\over2}m^2\phi^2$. Equation~(\ref{efold}) gives the integrated
expansion from $\phi_{\rm i}$ to $\phi_{\rm f}$ as
\begin{equation}
N\simeq {2\pi\over M_{4}^2}\left(\phi_{\rm i}^2-\phi_{\rm
f}^2\right)+{\pi^2m^2\over3 M_5^6}\left(\phi_{\rm i}^4-\phi_{\rm
f}^4\right)\,. \label{20}
\end{equation}
The new term on the right arising from the modified Freidmann
equation on the brane means that we always get more inflation for
a given initial inflaton value $\phi_{\rm i}$.

In the usual chaotic inflation scenario~\cite{chaotic} based on
Einstein gravity in four dimensions, the value of the inflaton
mass $m$ is required to be $\approx10^{13}$~GeV in order to obtain
the observed level of anisotropies in the cosmic microwave
background (see below). This corresponds to an energy scale
$\approx10^{16}$~GeV when the relevant scales left the Hubble
scale during inflation, but crucially also an inflaton field value
of order $3M_4$. Chaotic inflation has been criticised for
requiring super-Planckian field values to solve both the problems
of the standard background cosmology and lace the microwave
background with anisotropies of the observed magnitude. The
problem with super-Planckian field values is that one generically
expects non-renormalizable quantum corrections $\sim
(\phi/M_4)^n$, $n > 4$ to completely dominate the potential,
depriving one of control over the potential and typically
destroying the flatness of the potential required for inflation
(the $\eta$-problem~\cite{LytRio99}).

%%%%%%%%%%%%%%%%%%%%%%%%%%%
\begin{figure}
\epsfxsize=3.5in \epsffile{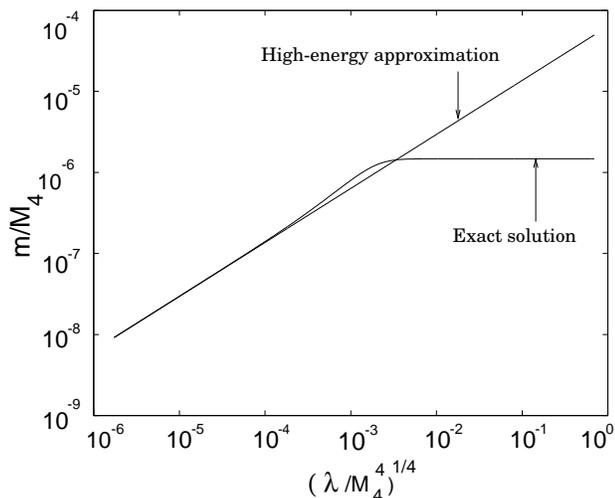} \caption{The scaling of
$m/M_{4}$ vs $(\lambda/M_4^4)^{1/4}$ to satisfy the COBE
constraints. The straight line is the approximation used in Eq.
(\ref{approx}), which at high energies is in excellent agreement
with the exact solution, evaluated numerically in slow-roll.}
\end{figure}
%%%%%%%%%%%%%%%%%%%%%%%%%%%

If the brane tension $\lambda$ is much below $10^{16}$~GeV,
corresponding to $M_5<10^{17}$~GeV, then the terms quadratic in
the energy density dominate the modified Friedmann equation. In
particular the condition for the end of inflation given in
Eq.~(\ref{endinf}) becomes $\dot\phi^2<{2\over5}V$. In the
slow-roll approximation [using Eqs.~(\ref{7}) and~(\ref{8})]
$\dot\phi\simeq-M_5^3/2\pi\phi$ and this yields
\begin{equation}
\phi_{\rm end}^4 \simeq {5 \over 4\pi^2}\left({M_5\over
m}\right)^2M_5^4 \,.
\end{equation}
In order to estimate the value of $\phi$ when scales corresponding
to large-angle anisotropies on the microwave background sky left
the Hubble scale during inflation, we take\footnote{The precise
value is dependent upon the actual energy scale during inflation
and the reheat temperature~\cite{LidLyt93}.  Our results are only
very weakly dependent upon the value of $N$ chosen.} $N_{\rm
cobe}\approx55$ in Eq.~(\ref{20}) and $\phi_{\rm f}=\phi_{\rm
end}$. The second term on the right of Eq.~(\ref{20}) dominates,
and we obtain
\begin{equation}
\label{phi55} \phi_{\rm cobe}^4 \approx {165\over\pi^2} \left({M_5
\over m}\right)^2M_5^4 \,.
\label{approx}
\end{equation}
Imposing the COBE normalization~\cite{blw} on the curvature
perturbations given by Eq.~(\ref{AS}) requires
\begin{equation}
A_{\sc s}\simeq \left({8\pi^2\over45}\right){m^4\phi_{\rm
cobe}^5\over
M_5^6} \approx 2\times10^{-5}\,.\label{22}
\end{equation}
Substituting in the value of $\phi_{\rm cobe}$ given by
Eq.~(\ref{phi55}) shows that in the limit of strong brane
corrections, observations require
\begin{equation}
m \approx 5\times 10^{-5}\, M_5\,,~~\phi_{\rm cobe}\approx 3\times
10^2\,M_5\,. \label{23}
\end{equation}
Thus for $M_5<10^{17}$~GeV, chaotic inflation can occur for field
values below the four-dimensional Planck scale, $\phi_{\rm
cobe}<M_4$, although still above the five-dimensional scale $M_5$.
The relation determined by COBE constraints for arbitrary brane
tension is shown in Fig. 1, together with the high-energy
approximation used above, which provides an excellent fit at low
brane tension relative to $M_4$.

\section{Conclusion}

In summary, we have found that slow-roll inflation is enhanced by the
modifications to the Friedmann equation in a cosmological scenario
where matter, including the inflaton field, is confined to a
three-dimensional brane, in five-dimensional Einstein gravity. This
enables the simplest chaotic inflation models, where the inflaton
potential is a polynomial in $\phi$, to inflate at field values below
the four-dimensional Planck scale.

We have calculated the expected amplitude of density perturbations
using the curvature perturbation $\zeta$ on uniform density
hypersurfaces, which we have argued will remain constant on very
large scales even in the presence of modifications to the Einstein
equations at high energies, so long as the perturbations are
adiabatic. Our calculations neglect the effect of gravitons in the
five-dimensional bulk which is always a consistent solution for
homogeneous matter fields~\cite{ShiMaeSas99}. However we note that
a full calculation should include the effect of back-reaction from
gravitational radiation in the bulk which might play an important
role for the high momentum wavemodes, possibly modifying the
amplitude of field fluctuations expected at Hubble-crossing.

Our results show that the additional friction term due to the
enhanced expansion at high energies drives the expected tilt of
the spectrum of density perturbations to zero, leading to the
canonical scale-invariant Harrison-Zel'dovich spectrum. The
modified dynamics alters the usual consistency relation between
the tilt of the gravitational wave spectrum and the ratio of
tensor to scalar perturbations expected in single-field slow-roll
inflation. At the same time the amplitude of tensor perturbations
is suppressed making an observational test of this prediction more
difficult. Conversely, the detection of a tensor signal would be
evidence against this scenario.

\section*{Acknowledgements}

The authors are grateful to David Lyth, Carlo Ungarelli and Kostya
Zloshchastiev for helpful discussions. DW is supported by the
Royal Society and IH is supported by the EPSRC.

\end{document}